\documentclass[aip,onecolumn]{revtex4}
\usepackage{graphicx}
\usepackage{epsfig}
\usepackage{bm}
\usepackage{amsmath}
\usepackage{amssymb}
\usepackage{wasysym}
\usepackage{epstopdf}
\usepackage{color}
\usepackage{xcolor}
\usepackage{epstopdf}
\usepackage{upgreek}

\usepackage[colorlinks=false, bookmarks=true]{hyperref}

\newcommand\ms{\nobreak\mbox{$\;$m\,s$^{-1}$}}
\newcommand\epsunit{\nobreak\mbox{$\;$m$^2$\,s$^{-3}$}}
\newcommand{\etal}{\emph{et al.}}

\begin{document}


\title[Lagrangian statistics of light particles]{Lagrangian statistics of light particles in turbulence} 
\author{Juli\'an Mart\'inez Mercado }
\altaffiliation[Present address: ]{Departamento de Fisica-FCFM, Universidad de Chile, Casilla 487-3, Santiago, Chile.}
\author{Vivek N. Prakash}
\email{v.n.prakash@utwente.nl}
\author{Yoshiyuki Tagawa}
\author{Chao Sun}
\email{c.sun@utwente.nl}
\author{Detlef Lohse}
\email{d.lohse@utwente.nl}

\affiliation{ 
Physics of Fluids Group, Faculty of Science and Technology, J. M. Burgers Centre for Fluid Dynamics, University of Twente, PO Box 217, 7500 AE Enschede, The Netherlands\\
(International Collaboration for Turbulence Research)}

\date{\today}

\begin{abstract}
We study the Lagrangian  velocity and acceleration statistics of light particles (micro-bubbles in water) in homogeneous isotropic turbulence. Micro-bubbles with a diameter $d_b=340$ $\mu$m  and Stokes number from 0.02 to 0.09 are dispersed in a turbulent water tunnel operated at Taylor-Reynolds numbers ($\mathrm{Re}_{\lambda}$) ranging from 160 to 265. We reconstruct the bubble trajectories by employing three-dimensional particle tracking velocimetry (PTV). It is found that the probability density functions  (PDFs) of the micro-bubble acceleration show a highly non-Gaussian behavior with flatness values in the range 23--30. The acceleration flatness values show an increasing trend with $\mathrm{Re}_{\lambda}$, consistent with previous experiments~\cite{Voth2002} and numerics~\cite{Ishihara2007}. These acceleration PDFs show a higher intermittency compared to tracers~\cite{Ayyala2008} and heavy particles~\cite{Ayyala2006} in wind tunnel experiments. In addition, the micro-bubble acceleration autocorrelation function decorrelates slower with increasing $\mathrm{Re}_{\lambda}$. We also compare our results with experiments in von K\'arm\'an flows and point-particle direct numerical simulations with periodic boundary conditions.
\end{abstract}


\maketitle

\section{\label{sec:level1}Introduction\protect\\ }
Multi-phase flows where the carrier fluid transports particles under turbulent conditions are ubiquitous. A thorough understanding of the dynamics of particles (light, neutral, or heavy) in turbulent flows is  therefore crucial. In most of these flows, the particles have a finite size and their density is different from that of the carrier fluid. Thus, the particle's dynamic behavior is expected to be different compared to neutral fluid tracers. 
The two relevant dimensionless parameters are the density ratio $\beta=3 \rho_f / (\rho_f+2\rho_p)$, where $\rho_f$ and $\rho_p$ are the fluid and particle density, and the Stokes number, which is the ratio of the particle's response time $\tau_p$ to the Kolmogorov  time scale  $\tau_{\eta}$, defined as St$=\tau_p/\tau_{\eta}=a^2/3\beta \nu \tau_{\eta}$, where $a$ is the particle radius and $\nu$ the kinematic viscosity of the carrier fluid.\\
The Lagrangian approach is naturally suited to study particles in turbulence and has recently attracted much attention \cite[see][]{Toschi2009}]. Pioneering Lagrangian particle tracking experiments in fully developed turbulence used silicon strip detectors to measure three-dimensional trajectories of tracer particles ($\beta$=1) with high spatial and temporal resolution in a \mbox{von K\'arm\'an} flow at high Taylor-Reynolds numbers, $\mathrm{Re}_{\lambda}$ up to 970 \cite[][]{Voth2002,Mordant2004,Mordant2004b}. The particle acceleration PDFs were found to be highly intermittent with flatness values around 55, and could be fitted with either stretched exponentials or log-normal distributions. The high intermittency of the fluid particle acceleration PDFs was also observed in numerical simulations \cite[]{Biferale2004,Mazzitelli2004,Toschi2005, Ishihara2007}. The normalized acceleration PDFs showed a weak dependence on the $\mathrm{Re}_{\lambda}$, and this was more prominently seen in the flatness values, which have been found to increase with $\mathrm{Re}_{\lambda}$ in both experiments~\cite{Voth2002} and numerics~\cite{Ishihara2007}. \\
More recent experimental investigations have focused on studying particles with different density than the carrier fluid~\cite{Qureshi2008,Volk2008a,Volk2008b,Gibert2010}. Heavy particles ($\beta=0$) in turbulence were studied using water droplets in wind tunnel experiments  at $\mathrm{Re}_{\lambda}$=250~\cite{Ayyala2006}. By following the particle motion with a moving camera, their trajectories were obtained in two-dimensions. It was observed that the normalized acceleration PDF was less intermittent than for tracers, with narrower tails. Numerical simulations~\cite{Bec2006a} have confirmed that the acceleration PDF of heavy particles is indeed slightly narrower than that of fluid tracers.\\
The dynamics  of light particles in turbulence ($\beta$=3) have been investigated both numerically  and experimentally.  DNS in the point-particle limit~\cite{Mazzitelli2004} showed very high intermittency in the PDFs of the individual forces acting on bubbles. A direct comparison of the statistics of light, neutral, and heavy particles was done by Volk \etal~\cite[][]{Volk2008a}, using data from both experiments and  point-particle DNS. In their experiments an extended laser Doppler velocimetry technique (extended LDV) was used to measure the particle velocity in a \mbox{von K\'arm\'an} flow at high $\mathrm{Re}_{\lambda}$=850. The experimental PDFs of the normalized acceleration  for light, neutral, and heavy particles did not reveal a visible difference within the experimental accuracy. In contrast, their numerical results showed that the acceleration PDF of light particles was more intermittent than that of tracers, and heavy particles showed less intermittency than tracers. They also found both in numerics and experiments, that  the acceleration  autocorrelation function of light particles decorrelates faster than those of neutrally buoyant and heavy particles.\\ 

In this work, we present an experimental study of the Lagrangian dynamics of light particles in turbulence. A majority of the previous Lagrangian particle tracking experiments focused mainly on tracer particles. Furthermore, grid-generated turbulence experiments in wind tunnels have dealt with either heavy or neutrally buoyant particles \cite{Ayyala2006,Ayyala2008,Qureshi2008}. Previous experiments  with bubbles \cite{Volk2008b} measured only one component of the velocity and acceleration in a \mbox{von K\'arm\'an} flow. Here, we provide results on the three components of the velocity, acceleration, and autocorrelation statistics of micro-bubbles ($\beta$=3) in homogeneous and isotropic turbulence. The micro-bubbles are dispersed in a turbulent water tunnel at moderate $\mathrm{Re}_{\lambda}$ (160---265), and their size is comparable to the Kolmogorov length scale. The micro-bubbles can respond to the small-scale fluctuations in the flow and hence the effects of finite-size are not important in this study.

The structure of this paper is as follows: in section \ref{sec:exp} we describe the experimental facility and the smoothing algorithm for the particle trajectories. The results are presented  in section \ref{sec:res}, followed by a conclusion and summary in section \ref{sec:con}.





\section{Experiments and Data Analysis}\label{sec:exp}

\subsection{\label{sec:level2}Experimental Setup}

 \begin{figure}
\centering
\includegraphics[width=0.45\textwidth]{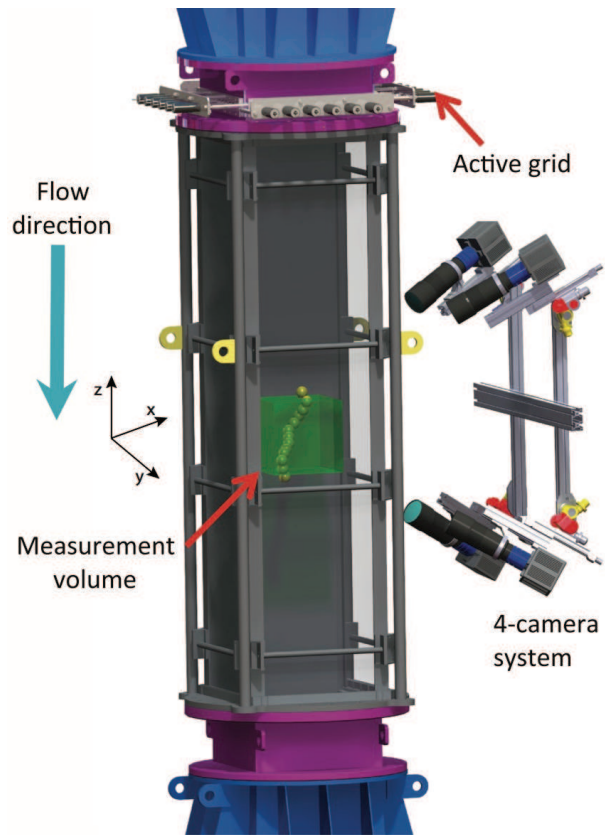}
\caption{The Twente Water Tunnel: an experimental facility for studying two-phase turbulent flows. The picture shows the measurement section and on top the active grid, which allows homogeneous and isotropic turbulent  flows upto $\mathrm{Re}_{\lambda}=300$, and the 4-camera particle tracking velocimetry (PTV) system to detect the positions of particles in three-dimensions. For illumination we use a high energy, high-repetition rate laser. Micro-bubbles with a diameter $\approx $340 $\upmu$m are generated above the active grid using a ceramic porous plate and are advected downwards into the measurement volume.} 
\label{fig:TWT}
\end{figure}

\begin{figure}
\centering
\includegraphics[width=0.55\textwidth]{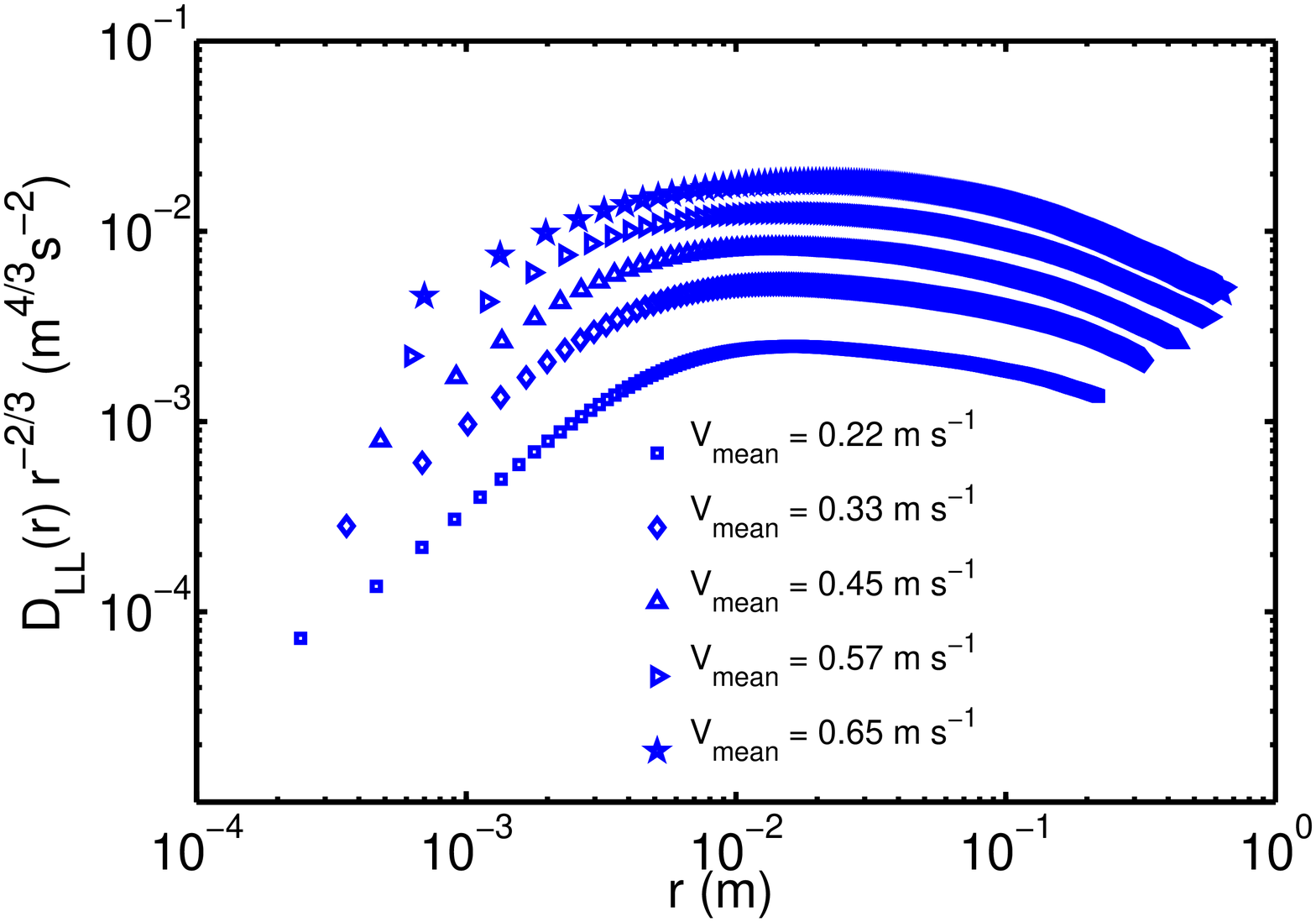}\\
\caption{Flow parameters characterization from hot-wire measurements. Compensated second-order longitudinal structure function $D_{LL}(r)$ calculated in order to estimate dissipation.} 
\label{fig:SF_HW}
\end{figure}

We conduct experiments in the Twente Water Tunnel, an 8 m long vertical water tunnel designed for studying two-phase flows (see Figure~\ref{fig:TWT}). By means of an active grid, nearly homogeneous and isotropic turbulence  with $\mathrm{Re}_{\lambda}$ up to 300 is achieved \cite[see][]{Rensen2005,Martinez2010}. A  measurement section  with dimensions 2$\times$0.45$\times$0.45 m$^3$ with three glass walls provides optical access for the three-dimensional PTV system.  Micro-bubbles  with a mean diameter $d_b=340 \pm 120$ $\mathrm{\mu}$m are generated by blowing pressurized air through a ceramic porous plate that is located  in the upper part of the water tunnel. These micro-bubbles are advected downwards by the flow and pass through the measurement section.\\ 
Our three-dimensional PTV system consists of four Photron PCI-1024 high-speed cameras that are synchronized with a high-energy (100 W), high-repetition rate (up to 10 kHz) Litron laser (LDY303HE). The four cameras are focused at the center of the test section on a $40\times40\times40$ mm$^3$ measurement volume that is  illuminated by expanding the laser beam with volume optics. The  arrangement of the cameras and laser is such that the four cameras receive forward scattered light from the micro-bubbles. We acquire images at 10,000 frames per second (fps) with a resolution of $256\times256$ pixels, resulting in a spatial resolution of about 156 $\upmu m$/ pixel.\\
 The $\mathrm{Re}_{\lambda}$ is varied by changing the mean flow speed of water in the tunnel. Table \ref{tab:flowCond} summarizes the flow properties for the various cases considered. The flow was characterized by measurements using a cylindrical hot-film  probe (Dantec R11) with a sampling rate of 10 kHz placed in the center of the imaged measurement volume. The dissipation rate $\epsilon$ was obtained from the Kolmogorov scaling for  the second-order longitudinal structure function $D_{LL}=C_2(\epsilon r)^{2/3}$, with $C_2=2.13$ \cite{Sreenivasan1995}. For each case of mean flow speed in the water tunnel, the dissipation rate is obtained from the value of the plateau region (see Figure~\ref{fig:SF_HW}), and other parameters follow. \\
 For the three-dimensional particle tracking, we use the open source code developed at the IfU-ETH group \cite[][]{Hoyer2005}. The error in the determination of the particle's position is within sub-pixel accuracy of 60 $\upmu m$, corresponding to the tolerance of epipolar matching in three dimensions. In this paper we focus on the Lagrangian statistics of micro-bubbles, but it is also possible to study particle clustering using the three-dimensional data (see~\cite{Tagawa2011}). Here, the raw particle trajectory is smoothed out with a polynomial fitting method (see section \ref{sec:smoothing}). Velocities and accelerations are obtained by differentiating the particle positions in the filtered trajectory. For the Lagrangian statistics shown in the results, the number of data points ($N_{data}$) used are larger than $4.5\times10^6$. 

\begin{table}
\caption{Summary of the flow parameters. $V_{mean}$: water mean flow speed, $\mathrm{Re}_{\lambda}=(15u_{rms}^4/\epsilon\nu)^{1/2}$: Taylor-Reynolds number, $u_{rms}$: mean velocity fluctuation, $\eta=(\nu^3/\epsilon)^{1/4}$ and $\tau_{\eta}=(\nu/\epsilon)^{1/2}$: are the Kolmogorov's length scale and time scale respectively, $L$: integral length scale of the flow, $\epsilon$: mean energy dissipation rate,  $St=\tau_p/\tau_{\eta}$: Stokes number,  and $N_{data}$: number of data points used to calculate the Lagrangian statistics.}
\begin{center}
 \begin{tabular}{|c|c|c|c|c|c|c|c|c|c|}
 \hline 
       V$_{mean}$&$Re_{\lambda}$&$u_{rms}$&$\eta$&$\tau_{\eta}$&$L$&$\lambda$&$\epsilon$&$St$ &$N_{data}$\\ 
       \ms& &\ms&$\mathrm{\mu}$m&ms&mm&mm&\epsunit& & \\ \hline 
             0.22&160&0.0161&400&160&64&9.9&39e-6&0.02&$5.5\cdot10^6$\\
        0.33&175&0.022&300&90&54&7.8&123e-6&0.04&$9.4\cdot10^6$\\
         0.45&195&0.027&250&65&56&7.0&237e-6&0.05&$8.3\cdot10^6$\\
          0.57&225&0.035&210&47&58&6.4&450e-6&0.07&$6.5\cdot10^6$\\
           0.65&265&0.043&180&35&70&6.0&786e-6&0.09&$4.5\cdot10^6$\\ \hline 
    \end{tabular}
    \end{center} \label{tab:flowCond}
\end{table}

\subsection{\label{sec:level2}Smoothing method for particle trajectories}\label{sec:smoothing}

Experimental errors in the determination of the particle positions are unavoidable, and obtaining the particles' velocity and acceleration through a time differentiation of their positions would be very sensitive to these experimental errors. Hence, a smoothing of the particle trajectory has to be carried out. This smoothing process is a trade-off between filtering out the experimental noise and retaining the turbulent features of the particle motion. Therefore, the smoothing parameters must be very carefully selected.

In the Lagrangian Particle Tracking community, two different methods have been used to smoothen particle trajectories in turbulent flows. One method consists of fitting the trajectory to a polynomial of second or higher order \cite[]{Voth2002, Luethi2005}, whereas the other method uses a Gaussian kernel~\cite{Mordant2004,Volk2011,Ayyala2006}. We tested both smoothing methods, and obtained very similar results. In the present paper, we only show results obtained by smoothing the micro-bubble trajectories with a third-order polynomial (also referred as the moving cubic spline method). 

\begin{figure}
  \centering
  \includegraphics[width=0.4\columnwidth]{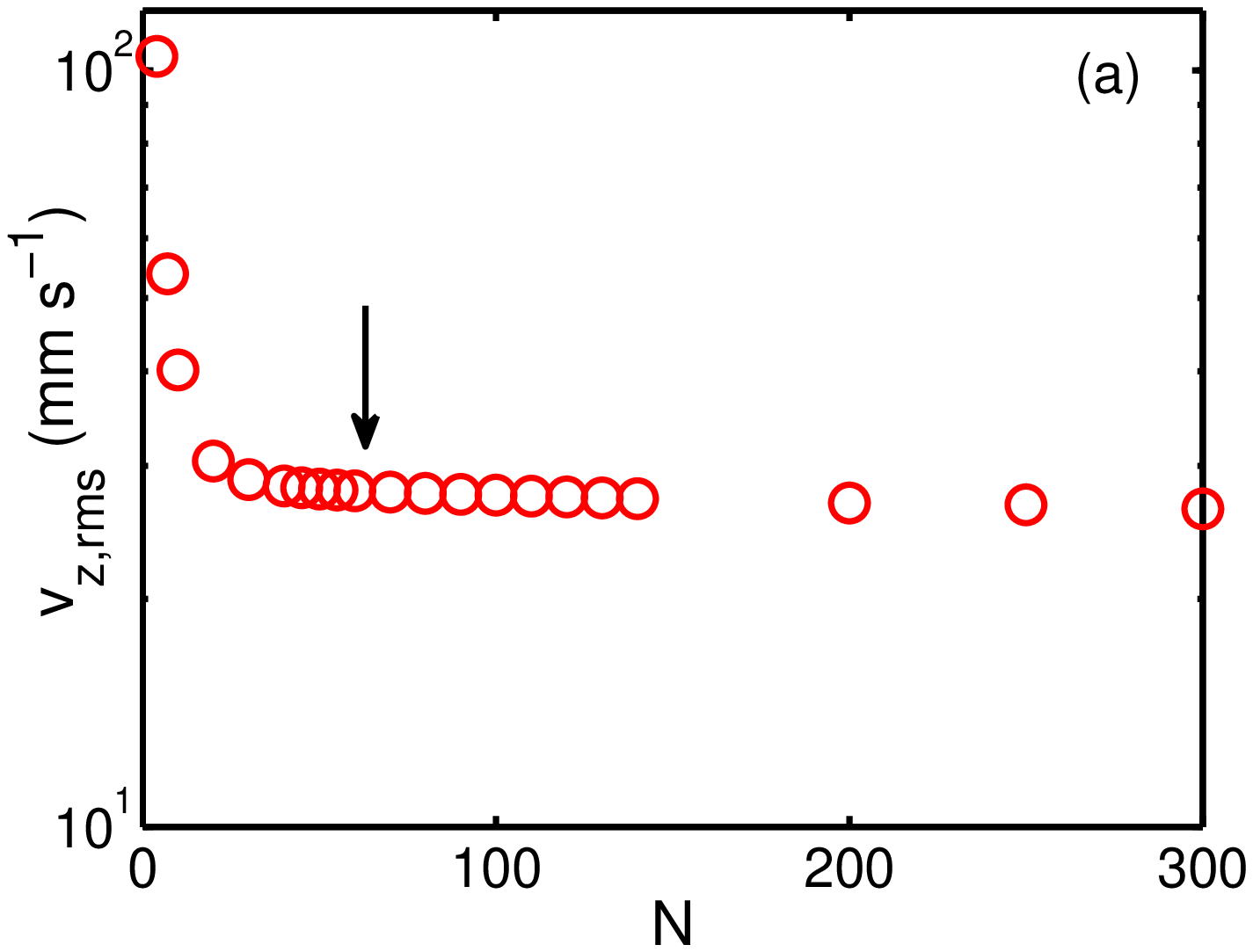}
  \includegraphics[width=0.41\columnwidth]{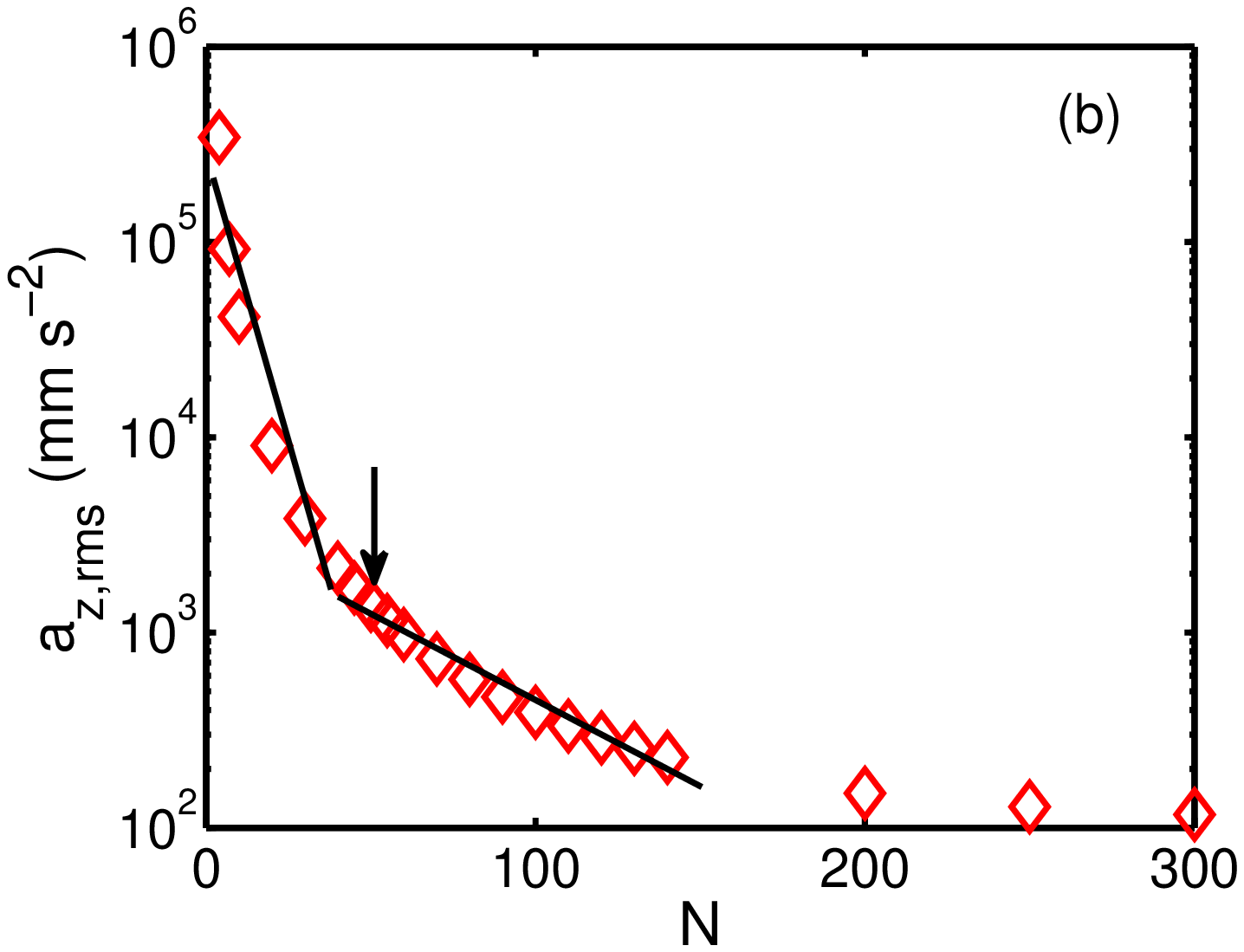}
\includegraphics[width=0.4\textwidth]{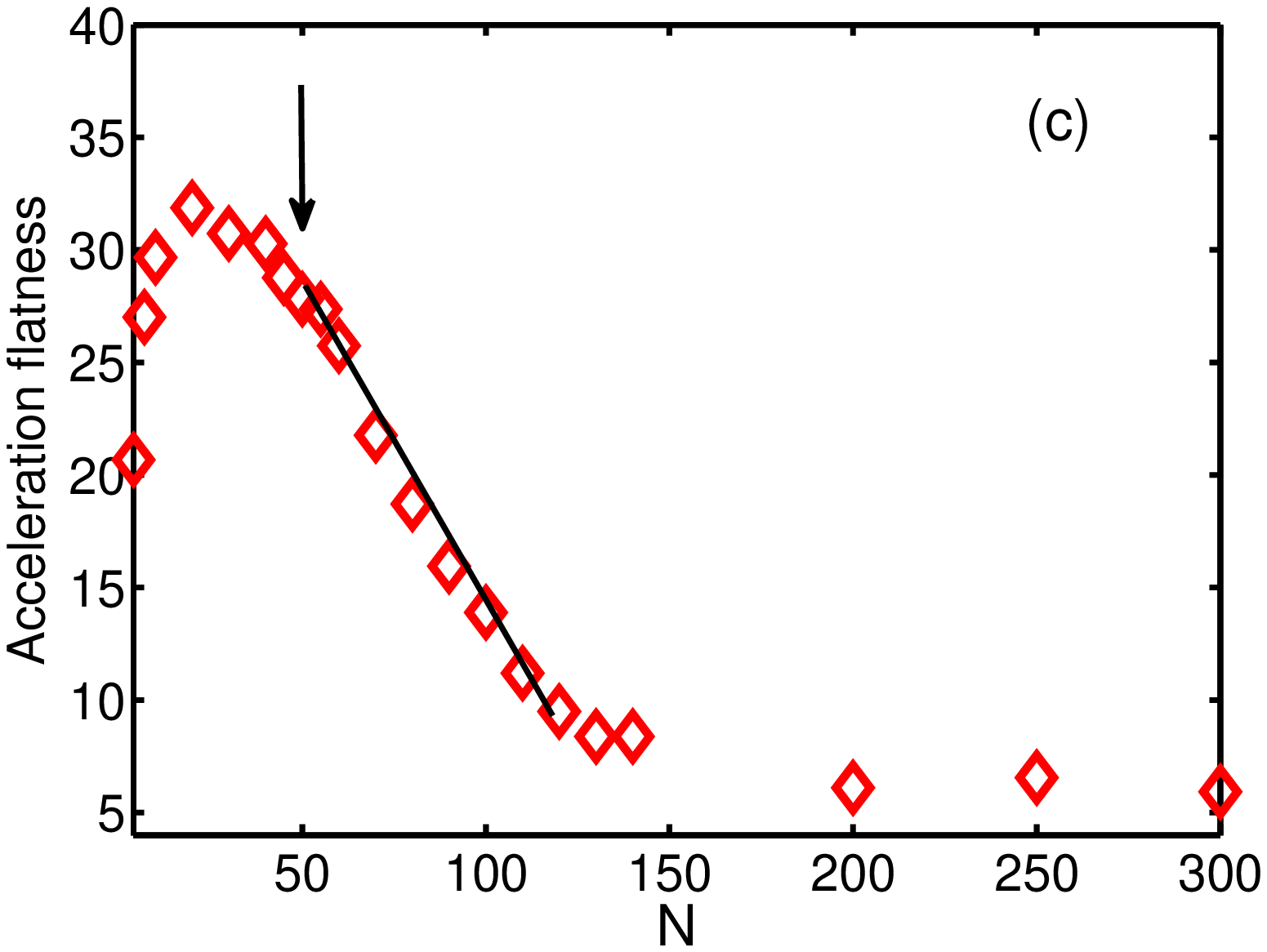}%
  \caption{r.m.s of the vertical component of the micro-bubble (a) velocity and (b) acceleration, and (c) the acceleration flatness at $\mathrm{Re}_{\lambda}$=195 as a function of $N$ for polynomial smoothing. The arrows in the figures indicate the chosen value ($N=50$) for the smoothing of the trajectories.} 
  \label{fig:rms}
\end{figure}

The entire signal of the trajectory $ x(t)$ is low-pass filtered by fitting a third-order polynomial. Using a fitting window with the particle positions from  $t-N dt$ until $t+N dt$, where $dt$ is the timestep, the filtered particle position at time $i$ is calculated as follows:
\begin{equation}
x_{i,f}(t)=c_{i,0}+c_{i,1}t+c_{i,2}t^2+c_{i,3}t^3.
\end{equation}
The Lagrangian velocity and acceleration are obtained by differentiating the particle trajectory:
\begin{equation}
u_{i,f}(t)=c_{i,1}+2c_{i,2}t+3c_{i,3}t^2,
\end{equation}
\begin{equation}
a_{i,f}(t)=2c_{i,2}+6c_{i,3}t.
\end{equation}
The parameter $N$ determines the length of the time window ($t-N dt$ , $t+N dt$), and has to be appropriately chosen to ensure that the time fitting window is smaller than the typical turbulent time scale. We explore the effect of $N$ on the r.m.s (root mean square) of the micro-bubble velocity (Figure~\ref{fig:rms}a) and acceleration (Figure~\ref{fig:rms}b) to find the optimum value for the case of $\mathrm{Re}_{\lambda}$=195. 
One can observe in Figure~\ref{fig:rms}a that the r.m.s of the velocity saturates at around $N=40$, for smaller values of $N$ there is an exponential rise owing to the noise. Since the acceleration is a second-order derived quantity, it is more sensitive to the choice of $N$ as observed in figure~\ref{fig:rms}b. Here, we can clearly distinguish two different regions: for small values of $N<30$ the r.m.s again increases exponentially due to the noise, while at large values of $N>100$ the $a_{rms}$ reduces considerably as an effect of the over-smoothing. For the data presented in this work we have chosen values of $N$ in the range 45---50, which correspond to normalized values of $N/ \mathrm{fps} \times \tau_{\eta}$ in the range 0.03---0.14.  It is important to point out that the normalized acceleration PDFs obtained by choosing $N$ in this range (45---50) are similar for each $\mathrm{Re}_{\lambda}$. The flatness of the acceleration PDF can also provide a measure to identify the optimal value of $N$ (as shown in Figure~\ref{fig:rms}c). We describe the details of the flatness calculation procedure in section~\ref{sec:flatness}. 
Here, in Figure~\ref{fig:rms}c, we see that our chosen optimal value of $N$ = 50 corresponds to the starting point in a region where the flatness values are decreasing with $N$ as a result of over-smoothing. At $N<10$, the decrease in flatness is an artificial effect arising from the noise. Hence, the optimal value of $N$ is chosen such that we do not over-smoothen the micro-bubble trajectories.


\section{\label{sec:res}Results}

\subsection{PDFs of micro-bubble velocity}
\begin{figure}[h!]
  \centering
\includegraphics[width=0.50\textwidth]{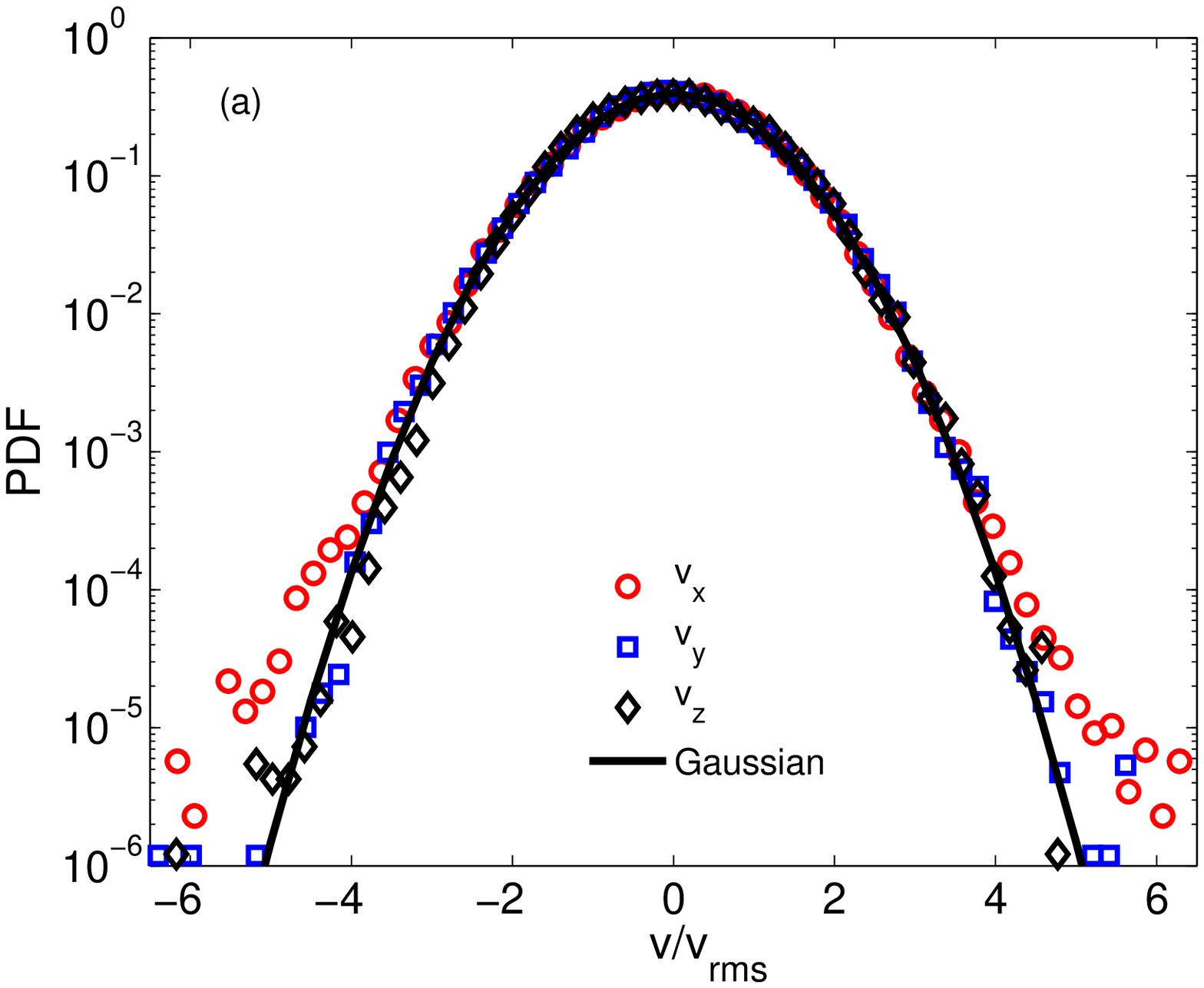}
\includegraphics[width=0.50\textwidth]{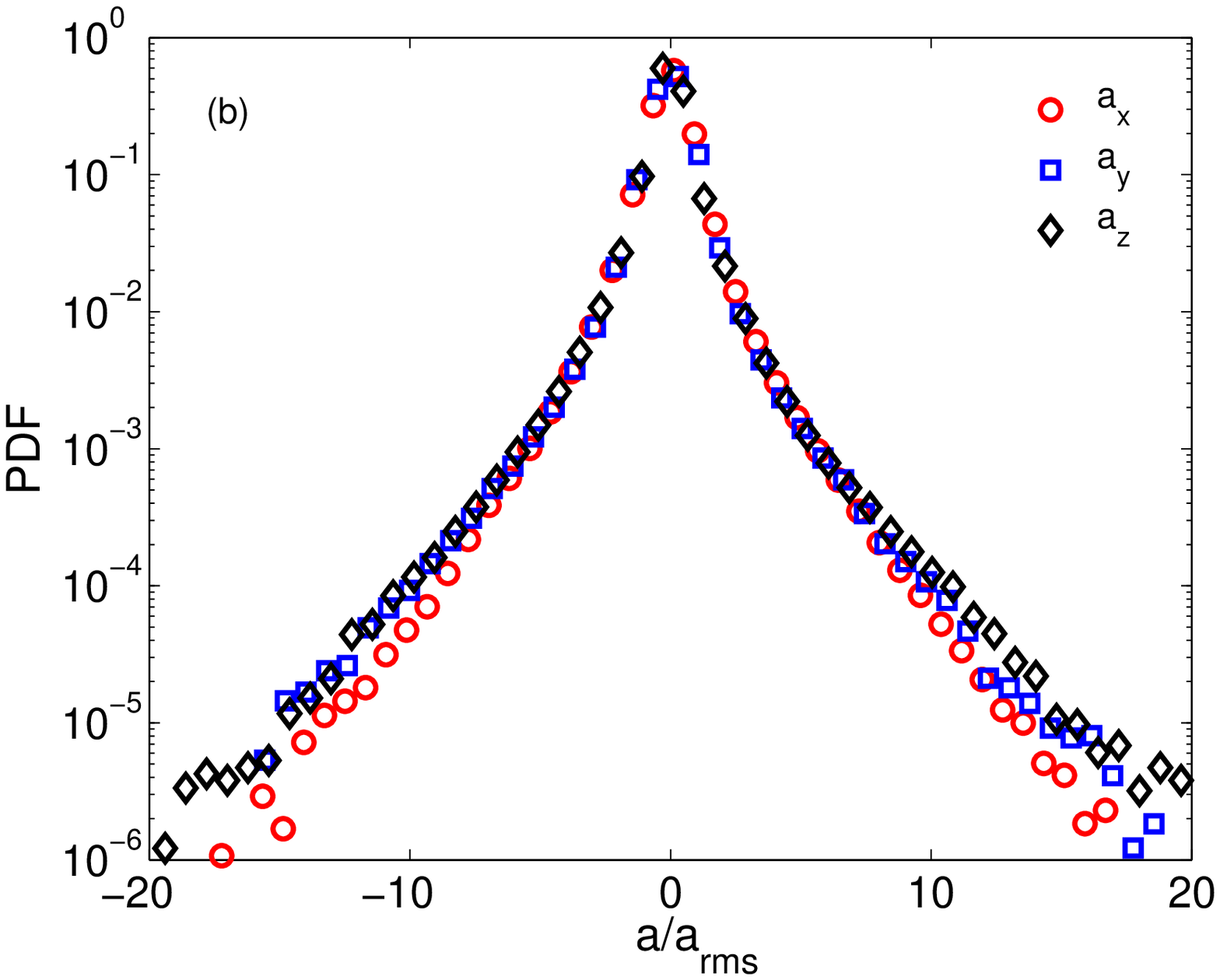}
  \caption{(a) PDFs of the three components of micro-bubble velocity at $\mathrm{Re}_{\lambda}=195$. The three velocity component distributions are nearly Gaussian compared to the solid line that represents a Gaussian distribution. (b) PDFs of the three components of the normalized micro-bubble acceleration at $\mathrm{Re}_{\lambda}=195$. The three components of the acceleration are strongly non-Gaussian, i.e. the tails of the distribution show high intermittency. }
  \label{fig:v_a_PDF}
\end{figure}

\begin{figure}
\centering
\includegraphics[width=0.45\textwidth]{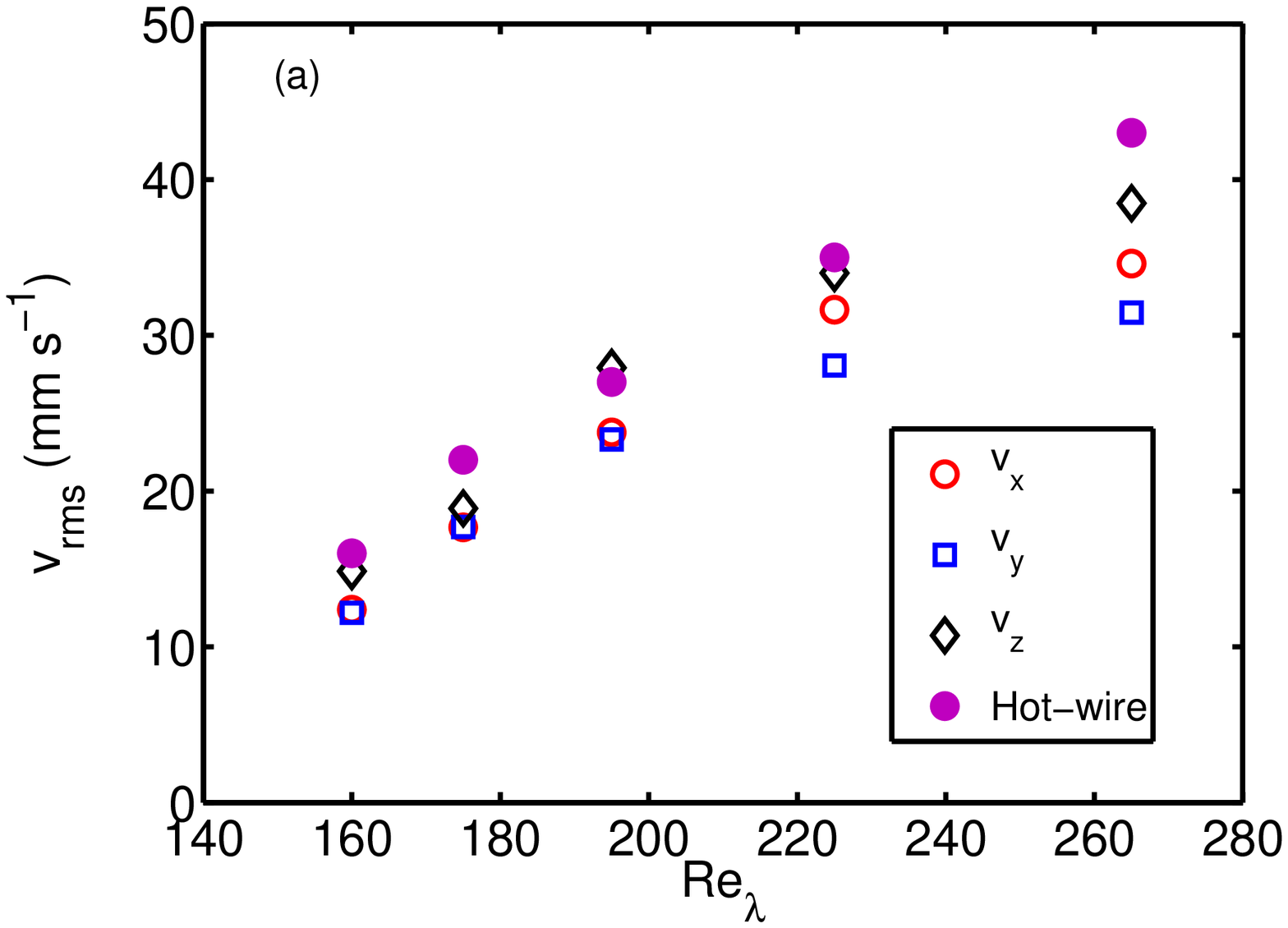}
\includegraphics[width=0.45\textwidth]{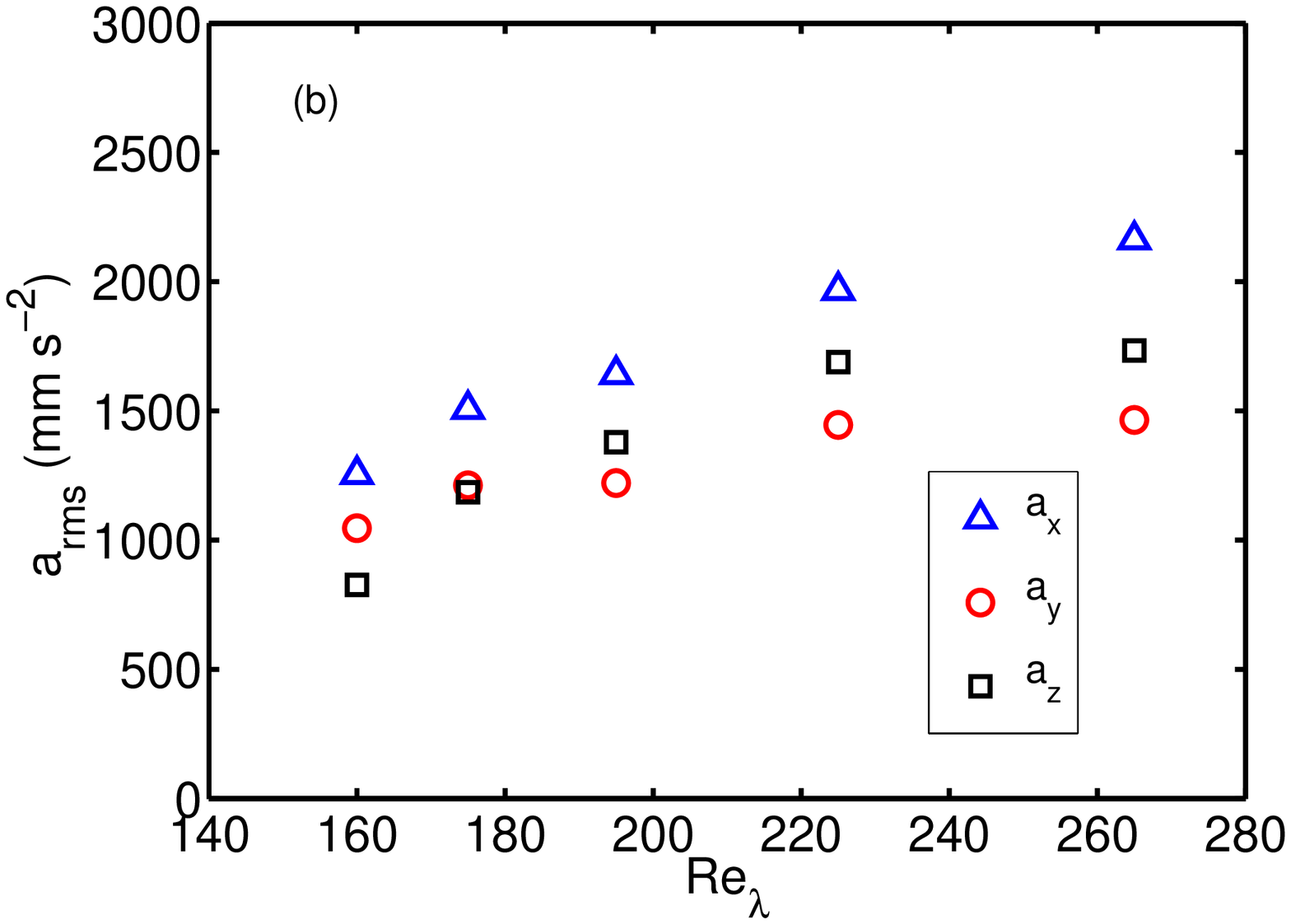}
\caption{(a) r.m.s values of the three components of the micro-bubble velocity for all $\mathrm{Re}_{\lambda}$, compared with the hot-wire probe measurements. (b) r.m.s values of the three components of the micro-bubble acceleration for all $\mathrm{Re}_{\lambda}$.}
\label{fig:urms}
\end{figure}

In this section we present results on the PDFs of micro-bubble velocity obtained by smoothing the raw trajectories. Figure~\ref{fig:v_a_PDF}a shows the PDF of the three components of the normalized micro-bubble velocity at $\mathrm{Re}_{\lambda}=195$. We observe that the velocity distributions of the three components closely follow a Gaussian profile. The flatness values $F$ of these PDFs for different $\mathrm{Re}_{\lambda}$ (see Table~\ref{tab:flatVel}) are close to that of a Gaussian distribution ($F=3$). Gaussian-type flatness values have also been reported for neutrally buoyant particles in turbulent von K\'arm\'an flows. Voth \etal~\cite{Voth2002} measured  velocity distributions close to Gaussian with flatness values in the range 2.8$-$3.2, and more recently, Volk \etal~\cite{Volk2011} obtained sub-Gaussian distributions with flatness around 2.4$-$2.6.
In Figure~\ref{fig:urms}a, we show a plot of the r.m.s values of the micro-bubble velocity versus $\mathrm{Re}_{\lambda}$. We observe an increasing trend with $\mathrm{Re}_{\lambda}$, and the three components are nearly isotropic. We also compare the r.m.s. values obtained from the hot-wire data and find a reasonable agreement with the 3D-PTV velocity measurements.

\begin{table}
\caption{flatness values of the distribution of micro-bubble velocities.}
\begin{center}
 \begin{tabular}{|c|c|c|c|}
 \hline 
 $Re_{\lambda}$ & $v_x$ & $v_y$ & $v_z$ \\ 
        \hline 
        160&$3.19 \pm 0.13$&$3.19 \pm 0.35$&$2.79 \pm 0.09$\\
        175&$3.09 \pm 0.05$&$3.08 \pm 0.33$&$2.96 \pm 0.14$\\
        195&$3.12 \pm 0.02$&$3.07 \pm 0.06$&$2.87 \pm 0.10$\\
        225&$3.04 \pm 0.04$&$3.04 \pm 0.09$&$3.25 \pm 0.16$\\
        265&$2.96 \pm  0.05$&$3.01 \pm 0.02$&$2.94\pm 0.06$\\ \hline 
    \end{tabular}
    \end{center} \label{tab:flatVel}
\end{table}

\subsection{PDFs of micro-bubble acceleration}

Contrary to the velocity PDFs, the micro-bubble acceleration PDFs normalized with the r.m.s ($\mathrm{a}/\mathrm{a}_{rms}$) exhibit a strong non-Gaussian behavior. Figure \ref{fig:v_a_PDF}b shows the PDFs for all the three components of the micro-bubble acceleration at  $\mathrm{Re}_{\lambda}=195$. We observe that the acceleration PDFs are highly intermittent with stretched tails that extend beyond $5\,\mathrm{a}_{rms}$, indicating that the probability of rare high acceleration events is much higher than for a Gaussian distribution. At this $\mathrm{Re}_{\lambda} (=195$), the acceleration is nearly isotropic as the PDFs of the three components show a good collapse atleast till $5\,\mathrm{a}_{rms}$. We have observed the same trend for the higher $\mathrm{Re}_{\lambda}$, whereas for smaller $\mathrm{Re}_{\lambda}$, the components of the acceleration ($\mathrm{a}_x$, $\mathrm{a}_y$) in the plane perpendicular to the mean flow direction are not yet isotropic. These components  have tails that are slightly narrower than the vertical component ($\mathrm{a}_z$). The flow in the Twente Water Tunnel is not fully isotropic, as has been discussed in Poorte $\&$ Biesheuvel~\cite{Poorte2002}. This slight anisotropy is visible in the PDFs, also in the $a_{x}$ component in Figure~\ref{fig:v_a_PDF}b. In Figure~\ref{fig:urms}b, we show the dependence of the acceleration r.m.s values on $\mathrm{Re}_{\lambda}$. Again, there is an increasing trend with $\mathrm{Re}_{\lambda}$ and a visible anisotropy in the three components. In the discussion that follows, we will only present results of the vertical component $z$ (mean flow direction) of the acceleration PDF.

\begin{figure}
 \centering
  \includegraphics[width=0.5\textwidth]{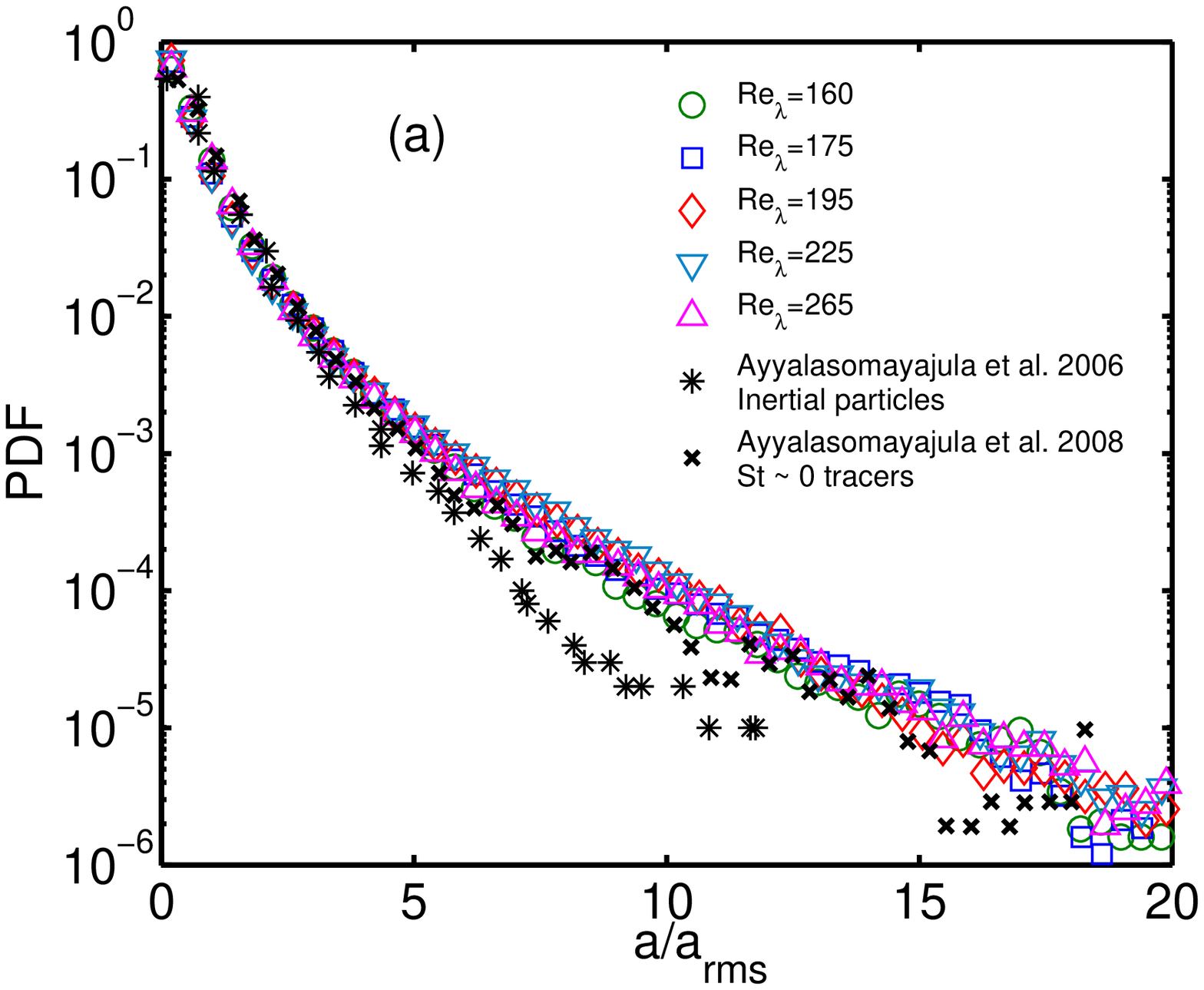}\\
  \includegraphics[width=0.5\textwidth]{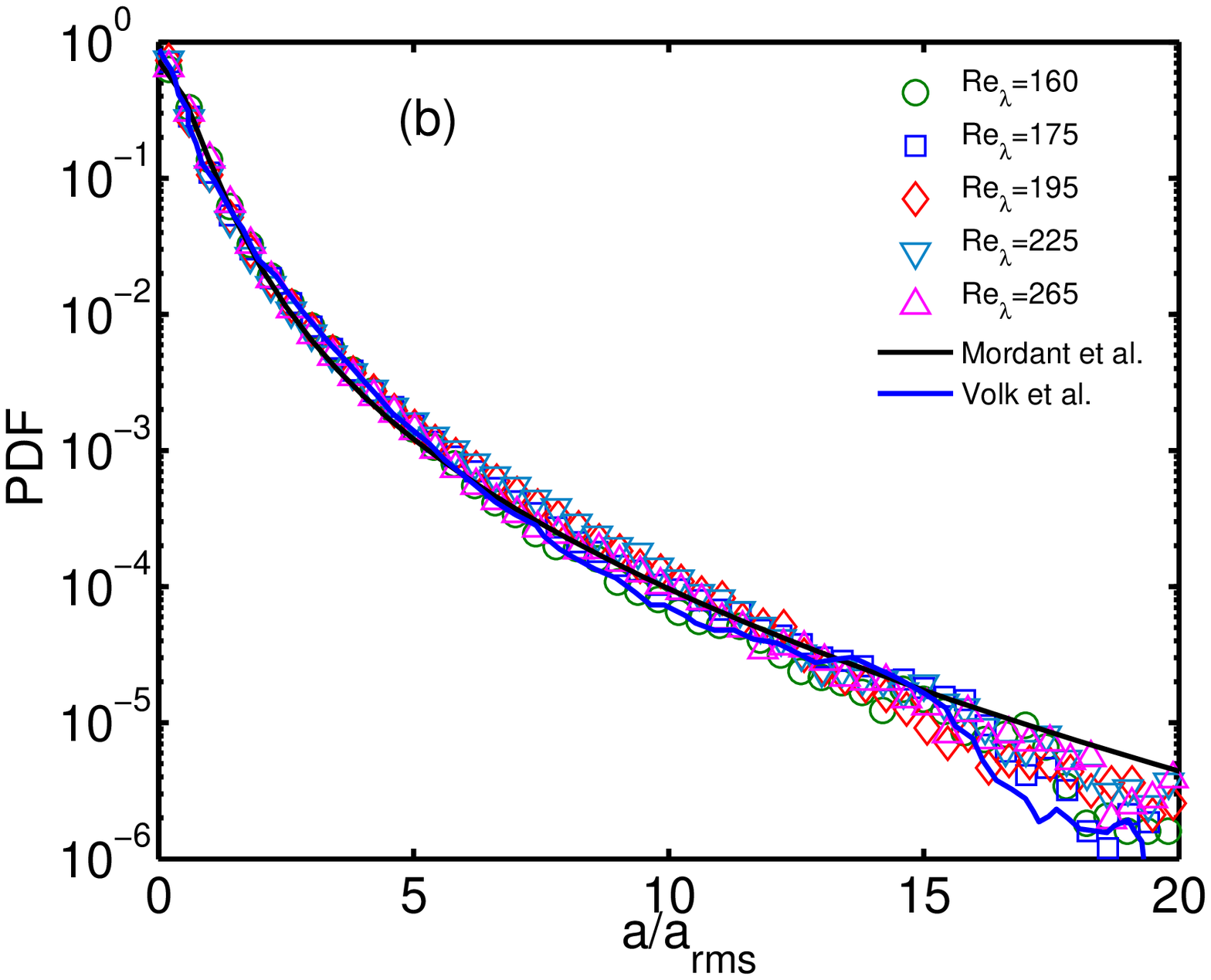}
  \includegraphics[width=0.5\textwidth]{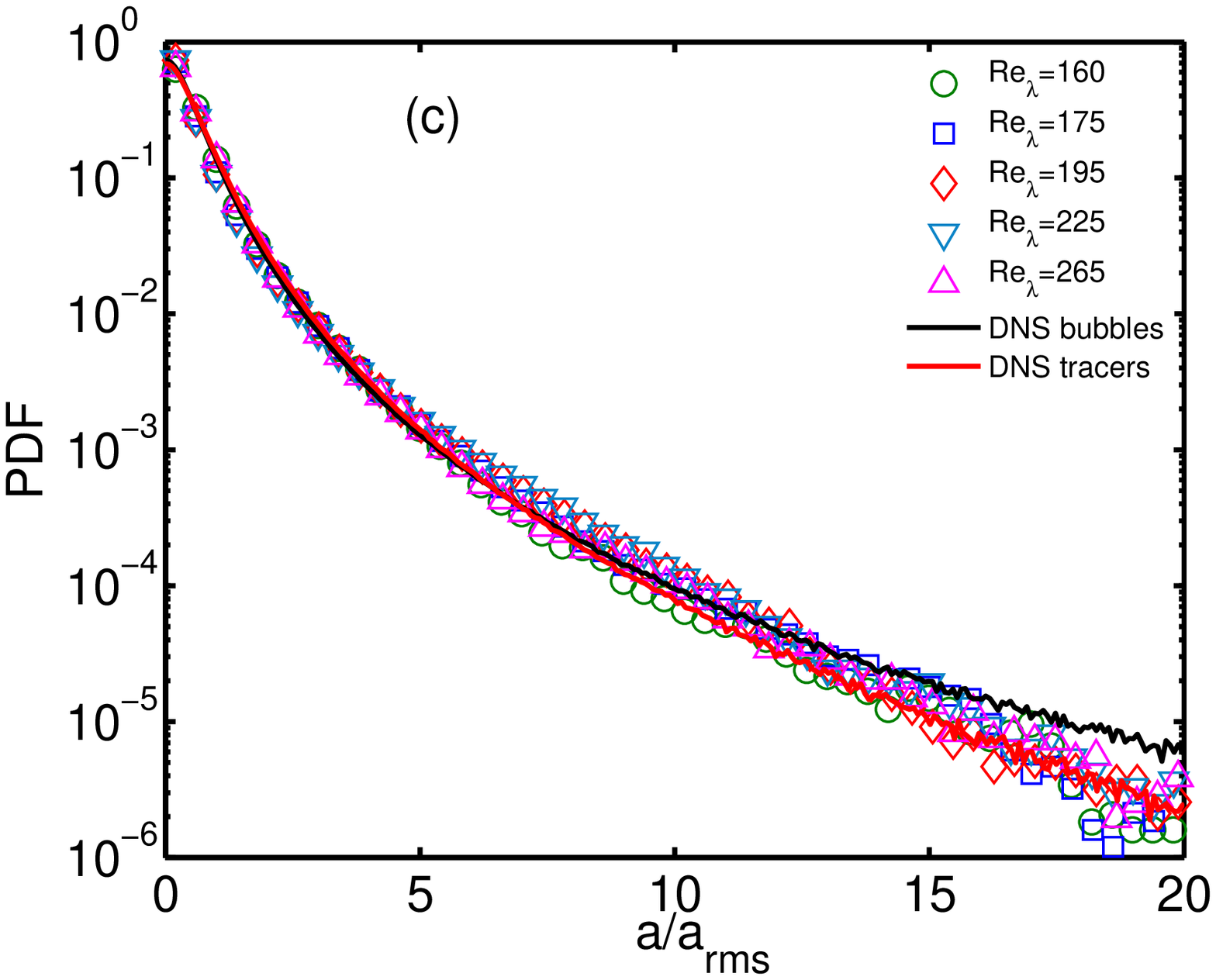}
  \caption{PDFs of the vertical component of the  normalized micro-bubble acceleration. (a) Comparison with experiments under similar flow conditions (grid-generated turbulence in a wind-tunnel) at $\mathrm{Re}_{\lambda}$ = 250. Our results are shown with open symbols; stars are heavy particles  \cite{Ayyala2006} and black crosses represent tracer particles \cite{Ayyala2008}. 
 (b) Comparison with von K\'arm\'an flow results: fit for tracers  at $\mathrm{Re}_{\lambda} = 140 - 690$ \cite{Voth2002,Mordant2004} is the black line;  bubbles at $\mathrm{Re}_{\lambda} = 850$ \cite{Volk2008a,Volk2008b} are shown with a blue line. (c) Comparison with DNS simulations for point particles at $\mathrm{Re}_{\lambda} = 180$ (from iCFDdatabase http://cfd.cineca.it): the red line indicates tracers and the black line bubbles.}
  \label{fig:az_all}
\end{figure}

\begin{figure}
\centering
\includegraphics[width=0.45\textwidth]{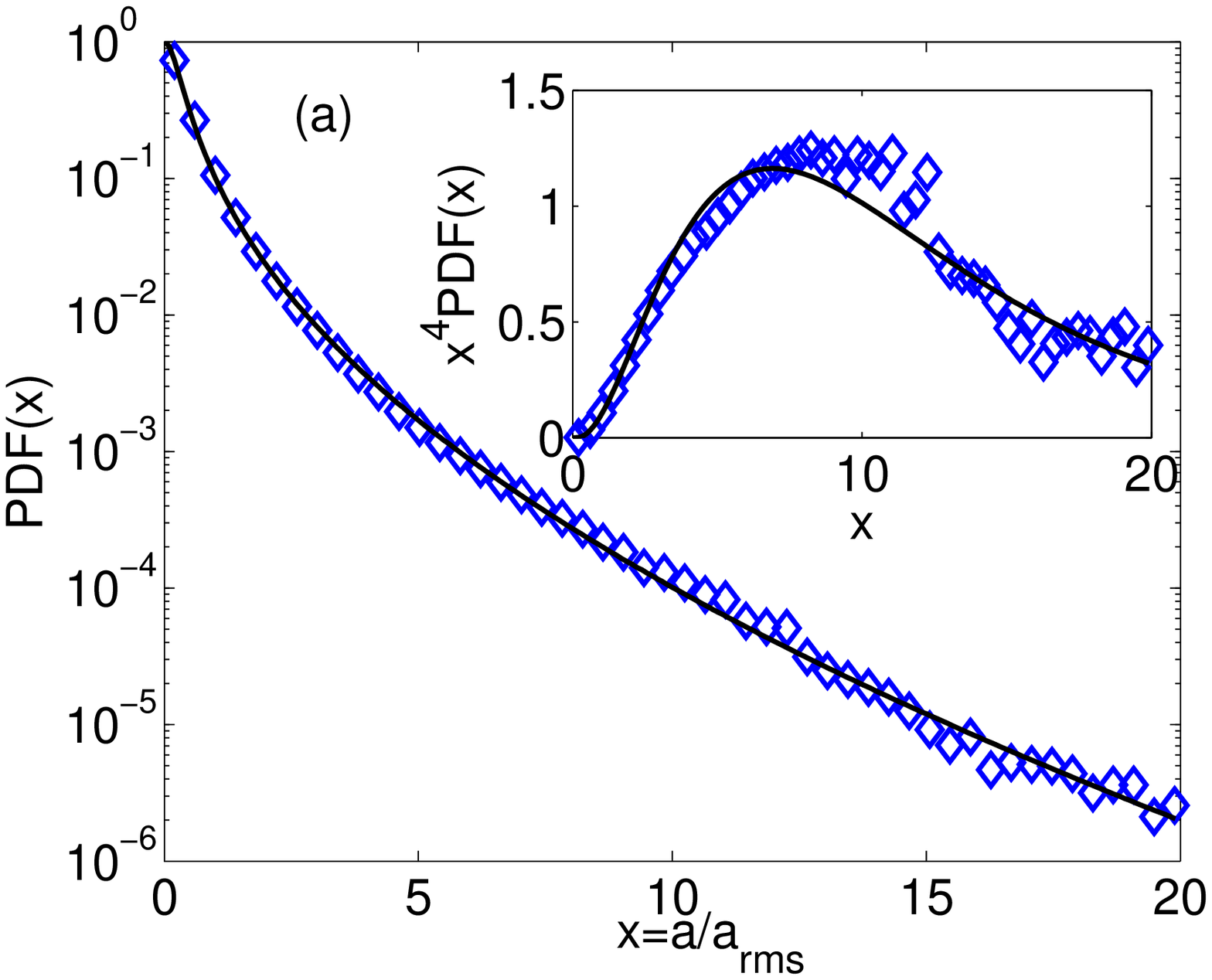}
\includegraphics[width=0.45\textwidth]{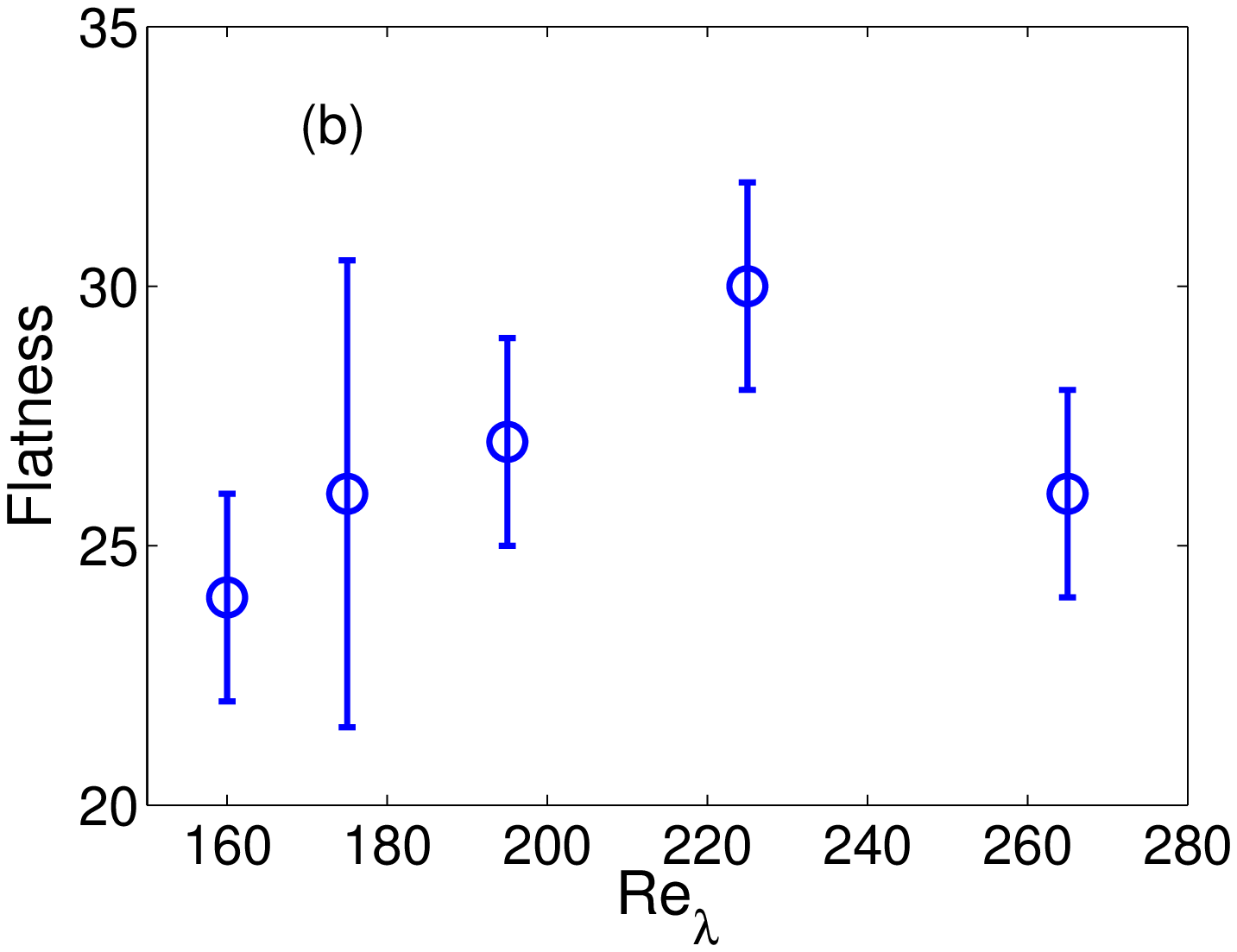}
\includegraphics[width=0.45\textwidth]{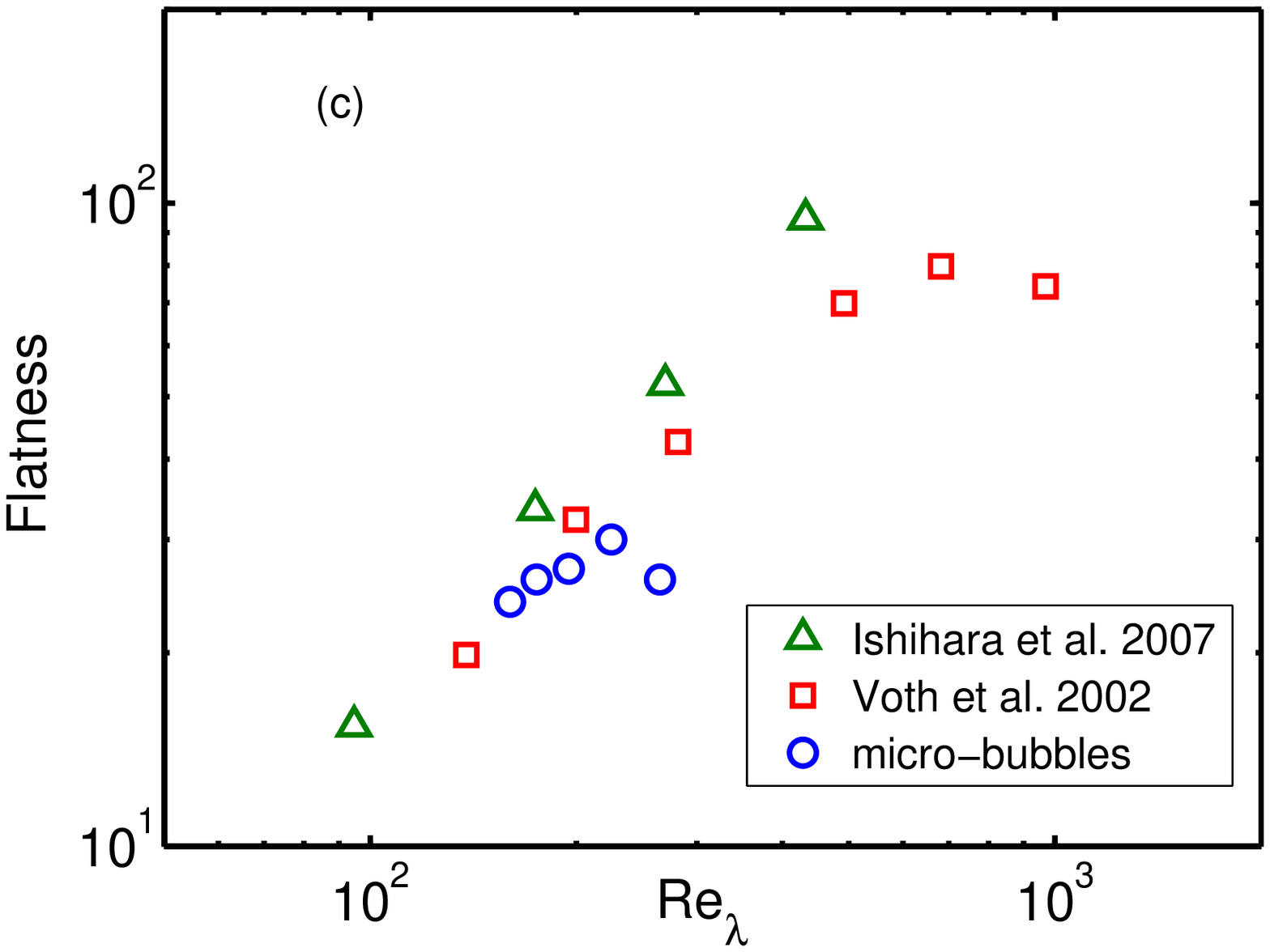}
 \caption{(a) PDF of the vertical component of the micro-bubble acceleration at $\mathrm{Re}_{\lambda}$=195. Open squares are the experimental data, solid line is the fitted stretched exponential function. The insert shows the plot of the fourth order moment $x^4P(x)$ for experimental data and fit. (b) The flatness value of the fitted PDFs of micro-bubble acceleration as a function of $\mathrm{Re}_{\lambda}$. (c) The flatness values versus the Reynolds number. Comparison with Voth \etal~\cite{Voth2002} and Ishihara \etal~\cite{Ishihara2007} reveals that the present micro-bubble trend agrees well with the data in the literature, at least till $Re_{\lambda} = 225$.}
 \label{fig:flatness_fig}
\end{figure}

In Figure \ref{fig:az_all}, we present the PDFs of the micro-bubble acceleration for all the $\mathrm{Re}_{\lambda}$ covered in the present study. In order to improve the statistics of the data, we take the absolute value of the acceleration vector. Although the r.m.s of acceleration increases with $\mathrm{Re}_{\lambda}$(Figure~\ref{fig:urms}b), the acceleration PDFs, normalized by the r.m.s, collapse on top of each other for all $\mathrm{Re}_{\lambda}$ (see Figure \ref{fig:az_all}). Here, we cannot see a clear dependence on $\mathrm{Re}_{\lambda}$, but the flatness of these acceleration PDFs better reveals the dependence, and will be discussed in section~\ref{sec:flatness}.

First, we compare our micro-bubble results with experimental data of heavy~\cite{Ayyala2006}  and tracer particles~\cite{Ayyala2008} under similar flow conditions (grid-generated turbulence with a mean flow). Ayyalasomayajula \etal~\cite{Ayyala2006} conducted experiments with heavy particles (water droplets) in a wind tunnel at St = 0.15 and $\mathrm{Re}_{\lambda}$ = 250, the corresponding results are shown in figure \ref{fig:az_all}a. Subsequently, measurements of tracer particles were also made in the same facility with St = 0.01 and $\mathrm{Re}_{\lambda}$ = 250 \cite{Ayyala2008} (also shown in figure \ref{fig:az_all}a). We observe that the present micro-bubble acceleration PDF shows a higher intermittency than heavy and tracer particles at similar $\mathrm{Re}_{\lambda}$ and St.\\
In figure \ref{fig:az_all}b, we compare the present data with experiments carried out in von K\'arm\'an flows. It is important to note the differences in the flow conditions between grid-generated turbulence and the turbulence generated in between counter-rotating disks. It is known that von K\'arm\'an flows have a large-scale anisotropy. Secondly, the confinement conditions are different. These differences could affect the Lagrangian dynamics~\cite{Qureshi2008}. Tracer particles in von K\'arm\'an flows at $\mathrm{Re}_{\lambda}$=140 to 690~\cite{Voth2002,Mordant2004} show a good agreement with our micro-bubble results for $|a_z/a_{z,rms}| \lesssim15$ (see the Mordant \etal~\cite{Mordant2004} fit in figure \ref{fig:az_all}b). But beyond this value our micro-bubble acceleration PDFs are slightly less intermittent. Furthermore, a comparison with the experiments of  Volk \etal~\cite{Volk2008b} who measured micro-bubble acceleration in von K\'arm\'an flow at $\mathrm{Re}_\lambda$ = 850 and St=1.85 is presented in figure~\ref{fig:az_all}b. Unexpectedly, there is a good agreement between the two experiments with very different $\mathrm{Re}_\lambda$, St, and flow conditions. This could just be a coincidence that both the results agree despite these differences. To arrive at a final conclusion on this issue, more systematic experiments need to be carried out in a wider $\mathrm{Re}_\lambda$-- St parameter space under our experimental conditions.\\ 
Figure~\ref{fig:az_all}c shows the comparison between the present micro-bubble acceleration results with the DNS data for point-like bubbles and tracers in homogeneous and isotropic turbulence at a similar $\mathrm{Re}_{\lambda}$=180 and St = 0.1 (data obtained from iCFDdatabase http://cfd.cineca.it). The simulations considered one-way coupled point particles within a periodic cubic box of size $L=2\pi$  and with a spatial resolution of $512^3$ (for further details on the simulation see~\cite{Bec2006a}). We observe that our experimental findings agree with both numerical bubbles and tracers when $|a_z/a_{z,rms}| \lesssim15$ within experimental error. The experimental PDF is closer to the numerical tracers for $|a_z/a_{z,rms}| \gtrsim15$. A possible reason for the better agreement between experimental micro-bubbles and DNS tracers could be the small St numbers O(0.01) in the present study. Another possible reason is the different flow conditions, in the experiments there is a strong mean flow which is absent in the numerics. In addition, several factors such as the lift force, buoyancy forces, and particle--particle interactions are ignored in the DNS \cite{Calzavarini2008b}. We emphasize again that in the present work, the micro-bubble size is comparable to Komogorov scale, hence, we do not expect the finite size effects \cite{Calzavarini2009} to play an important role.

\subsection{\label{sec:flatness}Flatness of the micro-bubble acceleration}

In order to quantify the intermittency of the acceleration PDFs, statistical convergence of the data is necessary. The number of data points needed for this convergence is crucial, and previous studies have shown that it should at least be $\approx$ O($10^6$) \cite[][]{Voth2002,Ayyala2006}. As shown in table \ref{tab:flowCond}, our measurements consist of at least 4.5 $\times$ 10$^6$ datapoints. 

The intermittency of the PDFs of the micro-bubble acceleration can be quantified by studying the flatness F:
\begin{equation}
\mathrm{flatness}=\mu_4/\sigma^4,
\end{equation}
where $\mu_4$ is the fourth moment and $\sigma$ the r.m.s of the distribution. The flatness being a fourth order moment is strongly determined by the tails of the distribution, and hence convergence of the PDFs is required. Even though the number of datapoints used to calculate the PDFs in the present work is larger than O($10^6$), full convergence has not yet been achieved to calculate directly the flatness from the distribution itself (the largest experimental datasets consist of $\approx$ O($10^8$) datapoints \cite{Mordant2004}, whereas for numerics this value can go up to $\approx$ O($10^9$) \cite[]{Biferale2004}. Consequently, we fit the experimental PDF to a stretched exponential distribution~\cite{Grossman1993,Mordant2004} defined as:
\begin{equation} \label{eq:se}
f(x)=C\exp\left (\frac{-x^2}{\alpha^2(1+|\frac{\beta x}{\alpha}|^\gamma)}\right),
\end{equation}
In equation \ref{eq:se}, $x=\mathrm{a}/\mathrm{a}_{rms}$ is the normalized micro-bubble acceleration, the fitting parameters are $\alpha$, $\beta$ and $\gamma$ and $C$ is a normalization constant. For this fitting procedure and in order to improve the convergence at the tails, we have taken the absolute value of the acceleration vector. 

Figure~\ref{fig:flatness_fig}a shows the result of the fitting for the micro-bubble acceleration PDF at $\mathrm{Re}_{\lambda}= 195$. The stretched exponential fits the experimental PDF quite well because the three fitting parameters enable a fine adjustment. In the inset of figure~\ref{fig:flatness_fig}a, we plot the fourth order moment $(\mathrm{a/a_{rms}})^4\mathrm{PDF(a/a_{rms})}$ for the experimental acceleration measurement along with the fitted curve. This type of curve allows for a good convergence test \cite{bel96}.  At the tails of the distribution, convergence is nearly achieved, and the fitted curve nicely sits on top of the experimental data.  We have observed  a similar behavior for the other measurements at different $\mathrm{Re}_{\lambda}$. 

Next, we calculate the flatness of the fitted acceleration PDFs as a function of $\mathrm{Re}_{\lambda}$, as shown in Figure~\ref{fig:flatness_fig}b. The flatness is determined directly from the fitted stretched exponential functional for all $\mathrm{Re}_{\lambda}$. The errorbars are obtained  by finding the difference between the flatness  values for half and the entire acceleration datapoints. Figure~\ref{fig:flatness_fig}c shows that the flatness values of the micro-bubble acceleration PDFs increase in the $\mathrm{Re}_{\lambda}$ range 160-225 consistent with the experimental results of Voth \etal \cite{Voth2002} for tracer particles and the numerical results of Ishihara \etal \cite{Ishihara2007} for fluid particles. For the highest $\mathrm{Re}_{\lambda}$, we have less statistics compared to the other cases as the mean flow speed is the fastest. This might be the reason for the decrease (underestimation) in the flatness value of our data point at $\mathrm{Re}_{\lambda}$ = 265. Clearly, from the collection of all data (Figure~\ref{fig:flatness_fig}c) one would not expect such a decrease.

\begin{figure}[h!]
\centering
\includegraphics[width=0.45\textwidth]{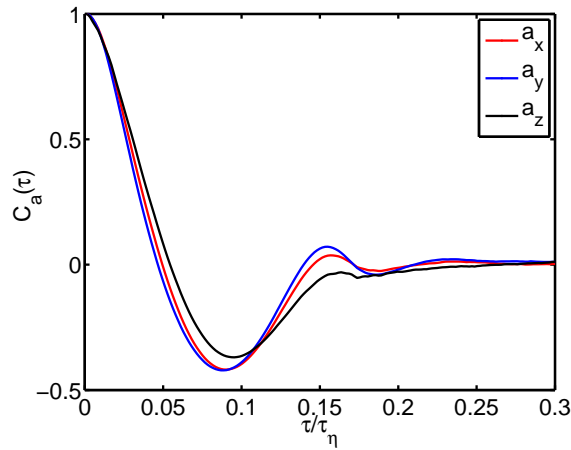}
\caption{Autocorrelation function of the three components of the micro-bubble acceleration at $\mathrm{Re}_{\lambda}$=195. The acceleration autocorrelation of the micro-bubbles is nearly isotropic. The time lag is normalized with the Kolmogorov time scale $\tau_{\eta}$.} 
\label{fig:aCorr}
\end{figure}

\begin{figure}[h!]
\centering
\includegraphics[width=0.5\textwidth]{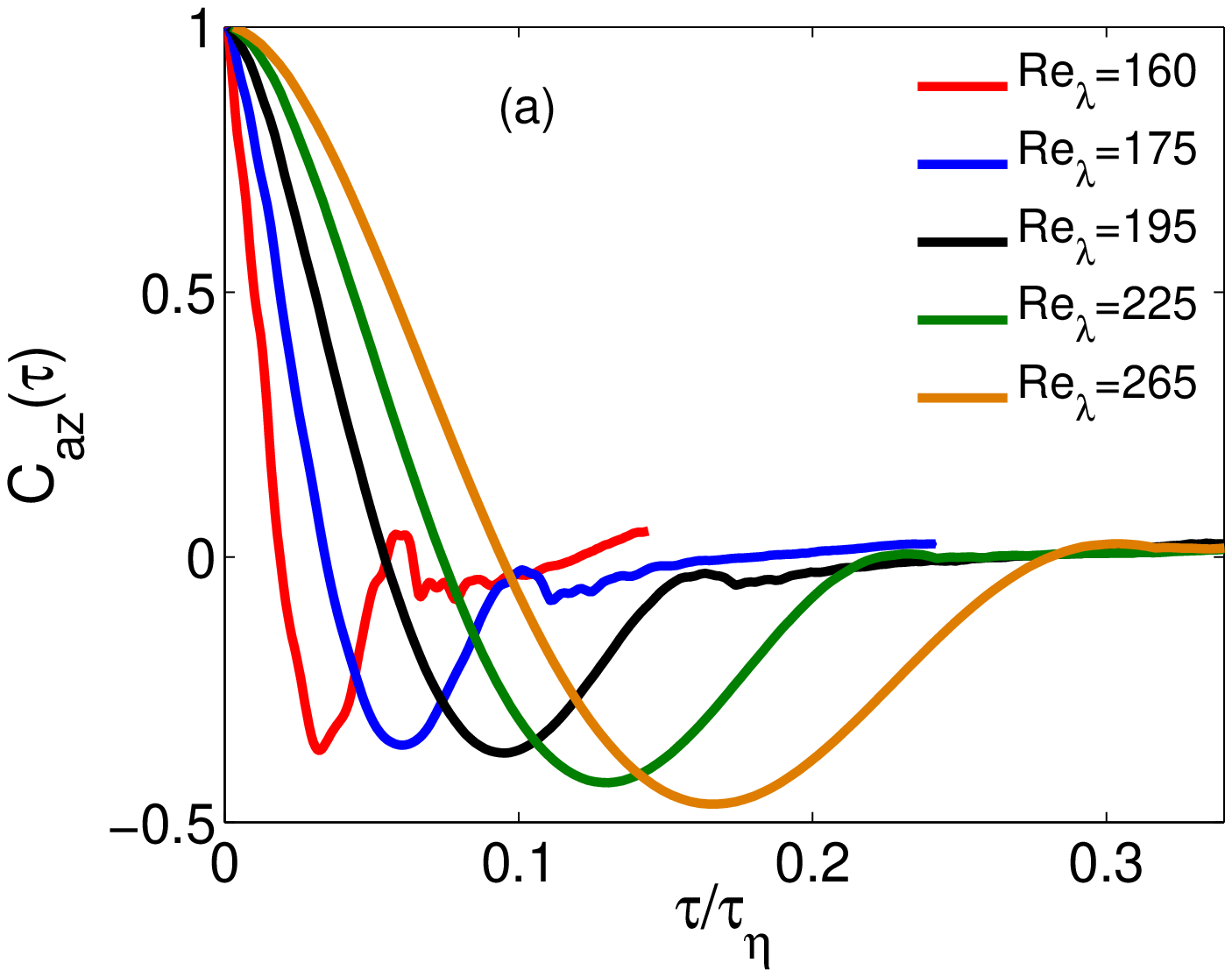}
\includegraphics[width=0.5\textwidth]{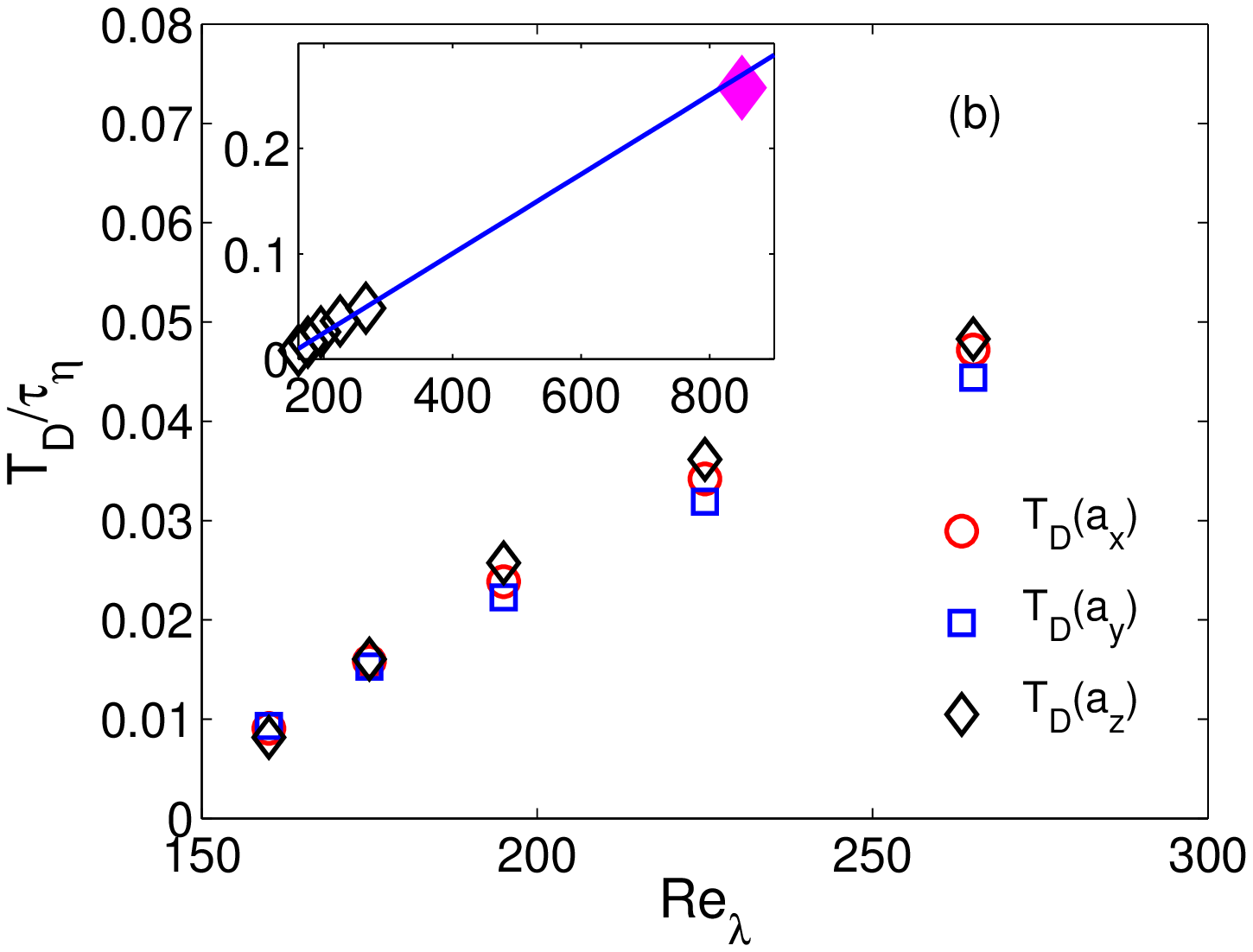}
\caption{(a) Autocorrelation function of  the vertical component of the micro-bubble acceleration for the different $\mathrm{Re}_{\lambda}$ measured. The correlation of the micro-bubble acceleration persists longer with increasing Reynolds number. (b) The decorrelation time $T_D/\tau_{\eta}$ of the autocorrelation function for the three components of the micro-bubble acceleration as a function of $\mathrm{Re}_{\lambda}$. The decorrelation time increases with the turbulent intensity. 
In the inset, we show also the result of Volk \etal~\cite{Volk2008a} at a very high $\mathrm{Re}_{\lambda}$=850  ({\color{-green}$\blacklozenge$}), their experimental point agrees with the trend of increasing decorrelation time with turbulent intensity. The linear fit obtained with our experimental data extrapolates a value of $T_D/\tau_{\eta}=0.27$ at $\mathrm{Re}_{\lambda}$=850, which is slightly higher than their experimental value of $T_D/\tau_{\eta}=0.25$.} 
\label{fig:azCorr_allMF}
\end{figure}

\subsection{Autocorrelation functions} \label{sec:autocorr}
We now present results on the Lagrangian autocorrelation function of the micro-bubble acceleration. In figure \ref{fig:aCorr} we compare the autocorrelation for the three components of the acceleration at $\mathrm{Re}_{\lambda}$=195, using a time lag normalized with $\tau_{\eta}$. We find that the three acceleration components correlate in a similar manner. This nearly isotropic behavior was also found for the other measurements at  different $\mathrm{Re}_{\lambda}$. 

Figure \ref{fig:azCorr_allMF}a shows the autocorrelation of $\mathrm{a}_z$ for different $Re_{\lambda}$. It is clear that the  microbubble's acceleration correlates for longer times as $\mathrm{Re}_{\lambda}$ increases. The acceleration autocorrelation function drops to zero rapidly, and the zero-crossing point has small values: $<$ $0.1 \tau_{\eta}$. Voth \etal~\cite{Voth2002} and Mordant \etal~\cite{Mordant2004b} reported values of around $2.2 \tau_{\eta}$ in their experiments with tracers in von K\'arm\'an flows at high turbulence intensities ($\mathrm{Re}_{\lambda}>690$). The value of $2.2 \tau_{\eta}$ was first found from DNS by Yeung \etal~\cite{Yeung1989}. Volk \etal~\cite{Volk2008a} performed both micro-bubble and tracer experiments in a von K\'arm\'an apparatus, and found that the decorrelation of the microbubbles is smaller than that of tracers at a given $\mathrm{Re}_{\lambda}$. We do not yet know the exact reason for the large disparity between $2.2 \tau_{\eta}$ for the fluid particles compared to $0.1 \tau_{\eta}$ for the present micro-bubbles. One possible reason is that our flow conditions are different as we have a strong mean flow. 

We study the time at which the autocorrelation function drops to zero for different $\mathrm{Re}_{\lambda}$ by defining the decorrelation time as:
\[
T_D=\int_0^{\tau_0} C_{a}(\tau)d\tau,  \quad \mathrm{with} \quad C_{a}(\tau_0)=0,
\]
 where $C_{a}$ is the acceleration autocorrelation function. $T_D$ represents the characteristic time for the evolution of the micro-bubble response to changes in the flow conditions. Figure \ref{fig:azCorr_allMF}b  shows the dependence of $T_D/\tau_{\eta}$ on $\mathrm{Re}_{\lambda}$ for the three components of the micro-bubble acceleration. We observe that $T_D/\tau_{\eta}$ increases with $\mathrm{Re}_{\lambda}$, and that the autocorrelation functions are nearly isotropic as evidenced by the very similar $T_D/\tau_{\eta}$ values for the different components. In the inset of figure \ref{fig:azCorr_allMF}b, the decorrelation time $T_D/\tau_{\eta}$ as obtained by Volk \etal~\cite{Volk2008a}  at $\mathrm{Re}_{\lambda}$=850 agrees well with our increasing trend of the decorrelation time with $\mathrm{Re}_{\lambda}$. We fit our experimental data of the decorrelation time of $\mathrm{a}_z$  to a linear relation $T_D/\tau_{\eta}=0.00038\mathrm{Re}_{\lambda}-0.051$ that is shown in the inset of figure \ref{fig:azCorr_allMF}b as a solid line.  Evaluating these relations at $\mathrm{Re}_{\lambda}$=850 gives  $T_D/\tau_{\eta}=0.27$,  which agrees well with their experimental value of $T_D/\tau_{\eta}=0.258$~\cite{Volk2008a}. More experiments are needed to fill the gap of $\mathrm{Re}_{\lambda}$. Very recently, Volk \etal~\cite{Volk2011} found an increase of $T_D/\tau_{\eta}$ with $\mathrm{Re}_{\lambda}$ for a fixed particle size, just as we find in our micro-bubble experiments.

\section{Conclusion} \label{sec:con}
We have presented experimental results on the Lagrangian statistics  of micro-bubble velocity and acceleration in homogeneous isotropic turbulence. Three-dimensional PTV was employed to obtain the micro-bubble trajectories. The PDFs of micro-bubble velocity closely follow a Gaussian distribution with flatness $F\approx3$, independent of $\mathrm{Re}_{\lambda}$. But the acceleration PDFs are highly non-Gaussian with intermittent tails. Although the acceleration PDFs themselves do not show a clear dependence on $\mathrm{Re}_{\lambda}$, the flatness values reveal a clear trend. We fit the experimental acceleration PDFs to a stretched exponential function and estimate the flatness based on the fitting. The flatness values were found to be in the range of 23--30 and show an increasing trend with $\mathrm{Re}_{\lambda}$. This trend is consistent with previous experimental \cite{Voth2002} and numerical \cite{Ishihara2007} results.

A comparison of our results with experiments in von K\'arm\'an flows \cite{Voth2002,Mordant2004,Volk2008a,Volk2008b} suggest that the present micro-bubble acceleration PDF is similar to tracers and bubbles (in von K\'arm\'an flows) for very different $Re_\lambda$. However, there are significant differences in the flow conditions between the two experimental systems. Therefore, it is more relevant to compare our results with previous investigations in similar flow conditions, i.e. grid-generated turbulence. We find that the acceleration PDFs of our micro-bubbles are more intermittent as compared to heavy and tracer particles in wind tunnel experiments at similar St and $\mathrm{Re}_{\lambda}$ \cite{Ayyala2006,Ayyala2008}.

Compared to DNS simulations in the point particle limit, our micro-bubble acceleration PDFs show a reasonable agreement with both numerical tracers and bubbles, but in the tails our data has a  better match with numerical tracers. One possible reason is the differences in flow conditions between the experiments and numerics. Another possibility is that the St in our experiments are small (0.02---0.09). It will be interesting to study the acceleration statistics of finite-sized bubbles at large St in turbulent flows.

We also calculate the autocorrelation function of the micro-bubble acceleration, and observed that the decorrelation time increases with $\mathrm{Re}_{\lambda}$. This finding is consistent with other experimental investigations \cite[][]{Volk2008a,Volk2011} at very high Reynolds number. More experimental data is needed to fill the gap of $\mathrm{Re}_{\lambda}$ in order to further study the scaling behavior.

\begin{acknowledgments}
We thank Enrico Calzavarini, Daniel Chehata  G\'omez, Beat L\"uthi (IfU-ETH), and Federico Toschi (TU/E) for useful discussions. This work was supported by the Foundation  for Fundamental Research on Matter (FOM) and industrial partners through the FOM-IPP Industrial Partnership Program: \emph{Fundamentals of heterogeneous bubbly flows}. We also acknowledge support from the COST Action MP0806: \emph{Particles in Turbulence}. The source of the DNS data was from the ICTR-iCFDdatabase (http://cfd.cineca.it). Finally, we thank Gert-Wim Bruggert, Martin Bos, and Bas Benschop for assistance with the experimental setup.

\end{acknowledgments}


\bibliographystyle{prsty_withtitle}
\bibliography{microbubble}


\end{document}